# Orbital−Energy Splitting in Anion Ordered Ruddlesden−Popper Halide Perovskites for Tunable Optoelectronic Applications

Gang Tang,[a, c] Vei Wang,[b] Yajun Zhang,[c] Philippe.Ghosez,[c] Jiawang Hong [a, *]

[a] School of Aerospace Engineering, Beijing Institute of Technology, Beijing, 100081, China

[b] Department of Applied Physics, Xi'an University of Technology, Xi'an, 710054, China

[c] Theoretical Materials Physics, Q-MAT, CESAM, University of Liège, B-4000 Liège, Belgium

**Corresponding Author**

*E-mail: hongjw@bit.edu.cn.

**ABSTRACT**

The electronic orbital characteristics at the band edges plays an important role in determining the electrical, optical and defect properties of perovskite photovoltaic materials. It is highly desirable to establish the relationship between the underlying atomic orbitals and the optoelectronic properties as a guide to maximize the photovoltaic performance. Here, using first-principles calculations and taking anion ordered Ruddlesden-Popper (RP) phase halide perovskites $Cs_{n+1}Ge_nI_{n+1}Cl_{2n}$ as an example, we demonstrate how to rationally optimize the optoelectronic properties (e.g., band gap, transition dipole matrix elements, carrier effective masses, band width) through a simple band structure parameter. Our results show that reducing the splitting energy $|\Delta c|$ of $p$ orbitals of B-site atom can effectively reduce the band gap and carrier effective masses while greatly improving the optical absorption in the visible region. Thereby, the orbital-property relationship with $\Delta c$ is well established through biaxial compressive strain. Finally, it is shown that this approach can be reasonably extended to several other non-cubic halide perovskites with similar $p$ orbitals characteristics at the conduction band edges. Therefore, we believe that our proposed orbital engineering approach provides atomic-level guidance for understanding and optimizing the device performance of layered perovskite solar cells.

**1. Introduction**

Halide perovskites-based solar cells have achieved more than 25% power conversion efficiency (PCE) due to their excellent optoelectronic properties.[1-6] Theoretical studies have revealed that the high-symmetry crystal structure plays a key role in determining the optoelectronic properties of the perovskite materials.[4, 7] Structural dimensionality, which directly relates to crystal symmetry, has been widely used as one metric to account for photovoltaic performances of absorbers.[5-6] For examples, perovskite solar cells based on the absorbers with three-dimensional (3D) structure generally exhibit higher PCE than the low-dimensional perovskites-based photovoltaic devices.[2, 8] While this concept successfully reveals the performance trends of most existing absorbers, it fails for some other materials. Emerging halide double perovskites (e.g., $Cs_2AgBiBr_6$)[9-10] are such an example of materials for which properties cannot be well explained from structural dimensionality. Although they are all structurally 3D, most of them have non-ideal optoelectronic properties including wide indirect band gap and large carrier effective masses, due to mismatched angular momentum of the frontier B-site atomic orbitals.[9] Compared with structural dimensionality, the recently proposed electronic dimensionality concept[11], which describes the connectivity of the atomic orbitals near the band edges, has been shown to help understanding better the reported optoelectronic properties such as bandgaps, carrier mobilities and defect levels. For example, the unfavorable optoelectronic properties of 3D $Cs_2AgBiBr_6$ can be well understood from its 0D electronic dimensionality. In $Cs_2AgBiBr_6$, both the valence band maximum (VBM, deriving Ag $4d$ and Br $5p$ orbitals) and conduction band minimum (CBM, deriving Bi $6p$ orbitals) cannot connect in three dimension because the $[AgBr_6]$ octahedra are isolated by the adjacent $[BiBr_6]$ octahedra.[11] Therefore, it is of crucial importance to explore the orbital-property relationship in halide perovskites in order to improve our understanding and design materials with better photovoltaic properties.

Recently, we introduced the concept of orbital engineering[12] to explore the orbital-property relationship in halide perovskites by manipulating the band-edge orbital characters or components. The realization of orbital engineering includes two aspects: (i) identifying the key orbital hybridization characters at band edges to screen and design novel functional materials and (ii) manipulating the orbital components of band edges to modulate and optimize the material properties. Regarding the first aspect, a typical example is the use of lone-pair *s* orbital-derived antibonding states at the valence band edge as a screening rule to search for promising optoelectronic materials.[3] Another example is the rational control the *d-p* orbital hybridization to achieve flat conduction band and flat valence band simultaneously in a 3D halide perovskite.[12] Regarding the second aspect, there are common means of regulating the orbital components near the band edges. A typical example is the minimization of the crystal field splitting of orbital energies to realize high orbital degeneracy from doping and biaxial strain, thereby discovering and designing the high-performance layered thermoelectric materials.[13] Nowadays, in high-performance perovskite solar cells, it is usual to adjust the band-edge orbital components through composition[14] or strain engineering[15] in order to achieve optimal electronic and optical properties. However, to our knowledge, there have been so far only few reports regarding the manipulation of the orbital-energy splitting to tune and optimize the optoelectronic properties of halide perovskites.

Recently, the family of low-dimensional halide perovskites was extended to all-inorganic naturally layered Ruddlesden−Popper (RP) phases[16-20] with ordered mixed halides, but their large band gaps ($E_g$ > 2.6 eV) partially limit their photovoltaic applications as absorbers. In this work, taking $Cs_{n+1}Ge_nI_{n+1}Cl_{2n}$ as an example, we demonstrate how to optimize the optoelectronic properties of layered (-like) perovskites through a simple scheme of orbital-energy splitting by using the first-principles method. Based on the calculated results, the orbital-property relationship with a simple band structure parameter is established as a guideline to optimize the optoelectronic properties (e.g., band gap, transition dipole matrix elements, carrier effective masses, band width) through biaxial compressive strain. Our results reveal that reducing the splitting energy $|\Delta c|$ of $p$ orbitals of B-site atom can effectively reduce the band gap and carrier effective masses, while greatly improving the optical absorption in the visible region. Furthermore, we show that this orbital engineering approach can be reasonably extended to several other non-cubic halide perovskites with similar $p$ orbitals characteristics at the conduction band edges. We believe that the orbital engineering strategy presented in our work provides valuable guidance for tuning and optimizing the photovoltaic performance of layered perovskite solar cells.

## 2. Results and discussions

Recently synthetized all-inorganic RP phase perovskites $A_2BI_2Cl_2$ (A = Rb, Cs; B = Pb, Sn, Cd) show a rare example of spontaneous anionic order in halide perovskites,[16-20] as shown in Figure 1a. They crystallize in the tetragonal $K_2NiF_4$-type structure with space group $I4/mmm$ (no. 139). No experimental evidence of transition to a lower symmetry phase has been reported yet. These compounds can be seen as naturally layered materials alternating, along the stacking $c$-direction, a perovskite $ABI_2Cl_2$ block in which $Cl^-$ ions occupy the equatorial sites (e.g., $ab$ in-plane) and $I^-$ ions occupy out-of-plane apical sites and a rocksalt $ACl_2$ block. They can equivalently be viewed as neutral ACl₄ planes alternating with two shifted neutral BI₂ planes. We choose $Cs_2GeI_2Cl_2$ as the representative compound to demonstrate our orbital engineering strategy. Since it has negligible effects on the electronic bands near the Fermi level of $Cs_2GeI_2Cl_2$ (see Figure S1), spin-orbit coupling (SOC) is not considered in this work to reduce the computational cost. The optimized structural parameters of $Cs_2GeI_2Cl_2$ are summarized in Table S1 and S2 in the Supporting Information. For the purpose of the illustration, we assume here that $Cs_2GeI_2Cl_2$ stays in the *I*4/*mmm* phase because high symmetry phase usually shows better optoelectronic properties than low symmetry phase[3, 7]. Although it is not guarantee to be true at room temperature and all strain conditions for this compound, this is supported by the fact that no experimental evidence of phase transition to lower symmetry has been reported yet in RP halide perovskite compounds and that such an similar assumption was successful in the search of new thermoelectrics[13].

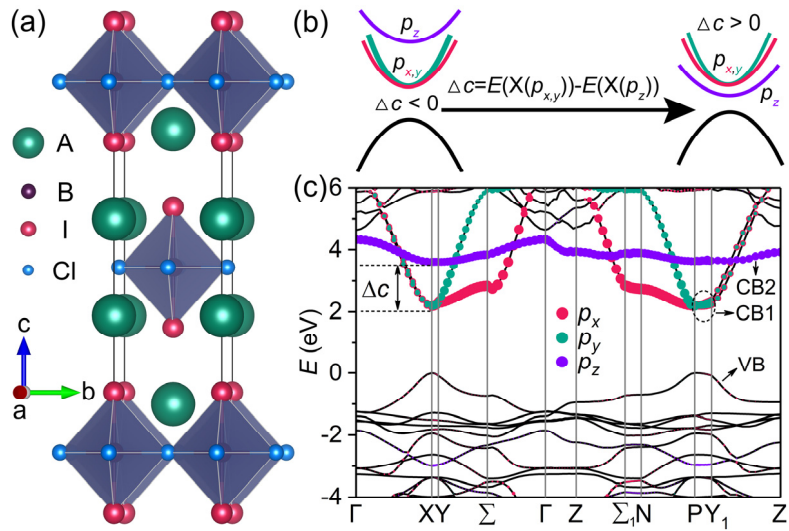

**Figure 1.** (a) Crystal structure of anion ordered Ruddlesden-Popper phase halide perovskites $A_2BI_2Cl_2$ (A = Rb, Cs; B = Pb, Sn, Cd, Ge). (b) Schematic diagram of orbital engineering by manipulating the relative energies of $p_x$, $p_y$ and $p_z$ orbitals at a specific high symmetry point (e.g., X). Nondegenerate band X($p_z$) and doubly degenerate band X($p_{x,y}$) are mainly composed of $p_z$ and $p_{x,y}$ orbitals from metal cations, respectively. $\Delta c$ is the splitting energy between $p_{x,y}$ and $p_z$ orbitals at the X point. (c) Orbital-projected band structure of representative compound $Cs_2GeI_2Cl_2$ with negative $\Delta c$ value. $p_{x,y}$ and $p_z$ orbitals of $Ge^{2+}$ cations are projected on the band structure. The corresponding first Brillouin zone is shown in Figure S2. Valence band (VB) is the topmost valence band, conduction band CB1 (CB2) is Ge $p_{x,y}$ ($p_z$) orbital-derived lower conduction band.

First, we check the thermodynamic stability of $Cs_2GeI_2Cl_2$ against chemical decomposition, as it has not yet been synthesized experimentally. The calculated decomposition enthalpies ($\Delta H_d$) along many possible pathways including redox and non-redox reactions are shown in Table S3. Two representative decomposition pathways with the lowest $\Delta H_d$ are as follows:

$$Cs_2GeI_2Cl_2 \rightarrow 2/3 CsGeCl_3 + 1/3 CsGeI_3 + CsI \qquad (1)$$

$$Cs_2GeI_2Cl_2 \rightarrow 2/3 CsGeCl_3 + 1/3 GeI_2 + 4/3 CsI \qquad (2)$$

Positive $\Delta H_d$ values (0.037 eV/f.u. for pathway (1) and 0.080 eV/f.u. for pathway (2)) along the above pathways indicate that $Cs_2GeI_2Cl_2$ is energetically favored. Further, we also verify the effects of exchange-correlation functional and mixed halogen configurations on the values of $\Delta H_d$. The calculated results (see Figure S3) further confirm that the $Cs_2GeI_2Cl_2$ structure displayed in Figure 1a is thermodynamically favorable.

Next, we will illustrate the basic idea of the orbital engineering strategy in layered perovskites (e.g., $Cs_2GeI_2Cl_2$), as shown in Figure 1b. In general, the optoelectronic properties of halide perovskites are strongly sensitive to the orbital characteristics at band edges.[3, 21] The conduction band edge is mainly dominated by the $p$ orbitals of B-site metal cations. In parent $CsBX_3$ cubic bulk perovskites ($Pm\bar{3}m$), the conduction band edge consists of three-fold degenerated $p$ orbitals at R point (1/2, 1/2, 1/2) of the cubic Brillouin zone.[21] For the layered perovskites (see Figure 1a), the conduction band edge is still made of $p$-states of the B-site metal atoms, but it is located at X point (1/2, 1/2, 0) of the tetragonal centred Brillouin zone and the $p_z$ orbital splits at higher energy than the $p_x$ and $p_y$ degenerated states (see Figure 1c). The energy difference between these two bands is defined as the splitting energy, namely, $\Delta c = E(X(p_{x,y})) - E(X(p_z))$. The $p$ orbitals of the B-site metal atoms contributing to CBM play an important role in the properties of the halide perovskites. Therefore, it is desirable to establish the relationship between the optoelectronic properties and the values of $\Delta c$ in layered halide perovskites, so that we can adjust the values of $\Delta c$ (e.g., from $\Delta c < 0$ eV to $\Delta c > 0$ eV) to optimize the corresponding electrical and optical properties.

Because the splitting energy $\Delta c$ is sensitive to change in the octahedral coordination environment, we tune the interlayer distance to manipulate the energy level of $p$ orbitals by applying the biaxial strain in the *ab* in-plane to achieve the tunability of $\Delta c$ (see Figure 2a). Interestingly, as can be seen from Figure 2a, there is a simple quadratic parabolic dependence between the biaxial strain $\varepsilon$ and the splitting energy $\Delta c$, which may be due to the non-linear change of Ge-I bond length along the out-of-plane direction (see Figure S4). Figure 2b-d show the dependence of the related optoelectronic properties on the values of $\Delta c$ for $Cs_2GeI_2Cl_2$. As shown in Figure 2b, the band gap decreases linearly with the decreasing $|\Delta c|$ value, when $\Delta c \leq 0$ eV. However, when $\Delta c$ becomes positive, meaning the energy levels of $p_{x,y}$ and $p_z$ orbitals are exchanged and $p_z$ orbital-derived band becomes the new CBM at X point, the band gap shows a large drop because the energy level of $p_z$ orbital is more sensitive to the interlayer hybridization compared to $p_{x,y}$ orbitals. The squares of the transition dipole matrix elements[22-24] $P^2$ at the X point are also calculated, revealing the transition probabilities between the topmost valence band (VB) and the lower conduction band (CB1 is Ge $p_{x,y}$-derived band and CB2 is the Ge $p_z$-derived band, see Figure 1c). Compared with the slight increase of $P^2_{VB \to CB2}$, $P^2_{VB \to CB1}$ shows a rapid increase with the increasing $\Delta c$ value, as shown in Figure 2b. The different increasing trends of $P^2$ will be explained in detail in the later sections.

Carrier effective masses ($m^*$) are another important descriptor to measure the electronic properties of photovoltaic materials. Due to the nature of the layered structure, $Cs_2GeI_2Cl_2$ exhibits strong anisotropic carrier transport characteristics. The less dispersive valence (conduction) bands along the out-of-plane direction (e.g., X-P direction) result in very large hole (electron) effective masses $m_h$ ($m_e$) (see Figure S5). Considering that the values of $m^*$ along the X-P direction tend to infinity, we only provide the evaluated results of $m_h$ and $m_e$ along the in-plane direction, as shown in Figure 2c. It can be seen that the hole effective masses $m_h$ gradually decrease with the increasing $\Delta c$ value, as reflected by the increased valence band widths $W_{VB}$ (see Figure 2d). The electron effective masses $m_e$ shows a slight decrease when $\Delta c < 0$ eV, while for $\Delta c \geq 0$ eV, it increases dramatically and then decreases. This is because the $p_z$ orbital-derived band becomes the new CBM and the band widths from Ge $p_z$-derived band ($w_{CB2}$) are significantly smaller than that of Ge $p_{x,y}$-derived band ($w_{CB1}$, see Figure 1d). The observed reduced in-plane $m^*$ and increased band widths can be well explained by the enhanced orbital hybridization in the in-plane direction under the compressive strain. Our results confirm that there is a certain correlation between the optoelectronic properties and the splitting energy $\Delta c$, which provides a valuable guide for experimentalists to optimize the related properties for photovoltaic applications.

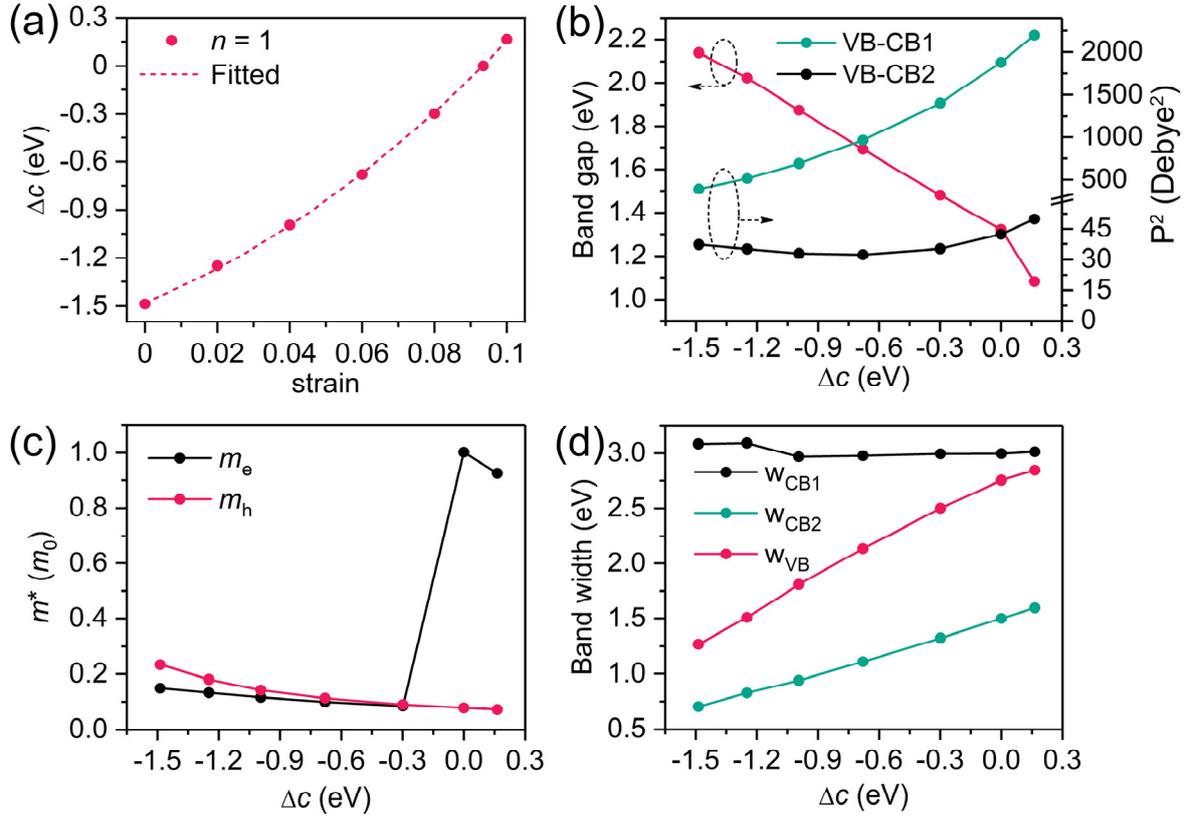

**Figure 2.** (a) Splitting energy $\Delta c$ as a function of biaxial compressive strain $\varepsilon$ in the representative Ruddlesden-Popper phase perovskite $Cs_{n+1}Ge_nI_{n+1}Cl_{2n}$ with $n$ = 1. Here biaxial strain $\varepsilon$ is defined as $|a - a_0|/a_0 \times 100\%$, where $a_0$ and $a$ are the in-plane lattice parameters with unstrained and strained states, respectively. The dependence of various optoelectronic properties on the value of $\Delta c$, such as (b) band gaps and transition dipole matrix elements, (c) carrier effective masses, and (d) band widths. Note that the symbols of VB, CB1 and CB2 are marked in Figure 1,

To better understand the evolution of optoelectronic properties with the change of $\Delta c$ value, we choose two special cases, one is the unstrained state ($\Delta c$ = -1.49 eV) and the other is the strained state with $\Delta c \approx 0$ eV (the $X(p_{x,y})$ and $X(p_z)$ bands are almost degenerate), and then calculated the band structure of $Cs_2GeI_2Cl_2$ at various *k* points along the high-symmetry directions, as shown in Figure 3. The corresponding Brillouin zones of primitive and conventional cells are displayed in Figure S1. We note that some recent literature reports[19, 25-26] employ conventional cells to calculate the band structure of $Cs_2BI_2Cl_2$ (B = Pb and Sn), which may not accurately reflect the true electronic properties due to the possible band folding effect. From Figure 3, it can be seen that the eigenvalue energy difference between X (0, 0, 0.5) and P (0.25, 0.25, 0.25) points is very small (< 100 meV) in this RP phase layered perovskites with the space group of *I*4/*mmm*. Comparing Figure 3a and 3b, we can find that as the value of |Δ *c*| decreases, the decrease in band gap is mainly attributed to the upward shift of valence band maximum (VBM). The projected density of states (PDOS) is further calculated to help to understand the underlying cause, as shown in Figure S6. The VBM is dominated by the antibonding states of Ge 4*s*, I $p_z$ and Cl $p_{x,y}$ orbitals. While the CBM is mainly composed of Ge 4*p* states. The observed significant upward shift of VBM results from that changes in the VBM energy level are more sensitive to lattice changes (e.g., in-plane biaxial compressive strain) than that of CBM energy level.

In addition, we also plot the sum of the squares of the transition dipole matrix elements $P^2$ along the different high-symmetry direction on the bottom of Figure 3a and 3b. In general, *P* is a complex vector quantity that includes the phase factors associated with a transition between an initial state *a* and a final state *b*, as denoted as $P_{a \to b}$:

$$P_{a \to b} = \langle \varphi_b | \mathbf{r} | \varphi_a \rangle = \frac{i\hbar}{(E_b - E_a)m} \langle \varphi_b | \mathbf{p} | \varphi_a \rangle \quad (3)$$

where $\varphi_a$ and $\varphi_b$ are energy eigenstates with energy $E_a$ and $E_b$, respectively; **p** and **r** represent the momentum operator and position operator, respectively; $\hbar$ is the reduced Planck constant; *m* is the mass of the electron. In Figure 3, it can be seen that with the decrease of |Δ *c*|

value, the significantly increased $P^2$ appears in the regions near X and P points, which is mainly contributed by the transition between VB and CB1 (see Figure 2b). According to the above definition, the influencing factors of $P^2$ can be divided into two parts. The first term is $1/(E_b-E_a)$, and the second term is $\langle\varphi_b|\mathbf{p}|\varphi_a\rangle$. Their different contributions to $P^2$ at the X point were calculated and shown in Figure S7. It can be seen that the obvious difference in the second term leads to the distinctly increasing trend of $P^2_{VB\to CB2}$ and $P^2_{VB\to CB1}$. We will see later it is $P^2$ that strongly contributes to the optical properties of $Cs_2GeI_2Cl_2$ and therefore tuning $\Delta c$ is an effective approach to optimize the optical properties.

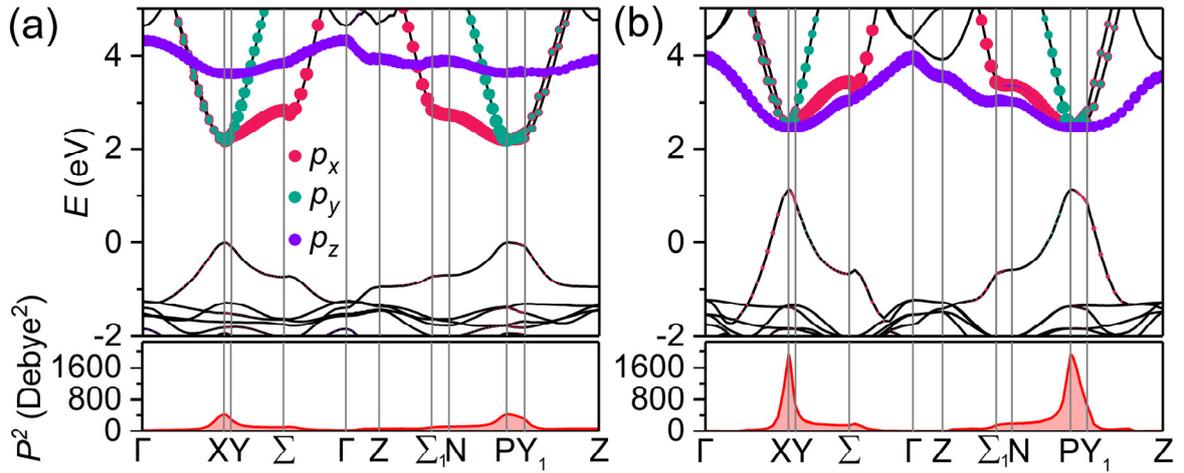

**Figure 3.** Orbital-projected band structure and corresponding transition dipole matrix elements for $Cs_2GeI_2Cl_2$ with (a) $\Delta c = -1.49$ eV and (b) $\Delta c = 0$ eV. $p_{x,y}$ and $p_z$ orbitals of $Ge^{2+}$ cations are projected on the band structure. Note that the energy scales are aligned with respect to the Cs 5$s$ states. $P^2$ are the sum of the squares of the transition dipole matrix elements between the topmost valence band (VB) and the lower conduction band (CB1 and CB2).

Next we will show how $P^2$ mostly contribute to the absorption of $Cs_2GeI_2Cl_2$. According to the Fermi golden rule[27], for a semiconductor, the optical absorption at photonic energy $\hbar\omega$ is directly correlated with:

$$\frac{2\pi}{\hbar}\int|\langle v|\mathcal{H}'|c\rangle|^2 \frac{2}{8\pi^3}\delta(E_c(\vec{k})-E_v(\vec{k})-\hbar\omega)d^3k \tag{4}$$

where $\mathcal{H}'$ is the perturbation associated with the light wave and $\langle v|\mathcal{H}'|c\rangle$ is the transition matrix from states in the VB ($v$) to states in the CB ($c$); $\delta$ is the Dirac delta function switching on this contribution when a transition occurs from one state to another, i.e., $E_c(\vec{k})-E_v(\vec{k})=\hbar\omega$. $E_c$ and $E_v$ show the conduction- and valence-band energies, respectively. The integration in the formula (4) is over the entire Brillouin zone. Considering that the matrix elements are almost constant within the Brillouin zone,[3-4, 24] for simplicity, the absorption formula is approximated as follows,

$$\frac{2\pi}{\hbar}|\langle v|\mathcal{H}'|c\rangle|^2 \int \frac{2}{8\pi^3}\delta(E_c(\vec{k})-E_v(\vec{k})-\hbar\omega)d^3k \tag{5}$$

where the second term is the joint density of states (JDOS) at energy $\hbar\omega$. According to this formula, it can be known that the optical absorption of a semiconductor is fundamentally determined by the transition matrix and JDOS. We then calculated the JDOS and optical absorption of $Cs_2GeI_2Cl_2$ for the above two cases ($\Delta c$ = -1.49 and 0 eV), as shown in Figure 4. From Figure 4a, we can see that the JDOS is very close in both cases ($\Delta c$ = -1.49 and 0 eV). As shown in Figure 4b, when the value of $\Delta c$ changes from -1.49 to 0 eV, the optical absorption of $Cs_2GeI_2Cl_2$ increases markedly in the visible region, indicating the optical properties are improved significantly by tuning $\Delta c$. More specifically, for the $\Delta c$ value of -1.49 eV, it takes about 0.7 eV for the absorption coefficient to rise to $10^3$ cm$^{-1}$, as indicated by the dotted box in Figure 4b. However, for the $\Delta c$ value of 0 eV, the energy range for the absorption coefficient to rise to the same level takes less than 0.2 eV.

According to above analysis, we can conclude that the increase in the optical absorption is mainly ascribed to the enhancement of $P^2$ near the band edges at the X and P points (see Figure 3).

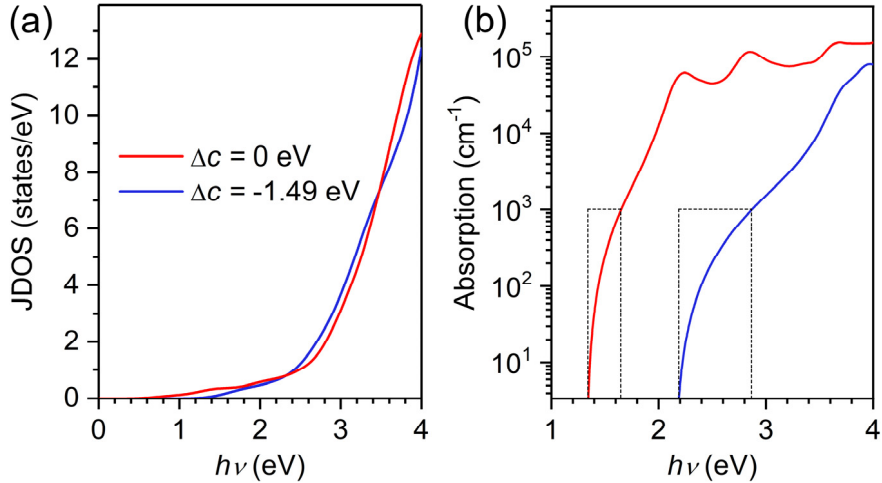

**Figure 4.** Calculated optical properties for $Cs_2GeI_2Cl_2$: (a) joint density of states and (b) absorption coefficient.

We demonstrate that our orbital engineering strategy work well for the layered perovskite $Cs_{n+1}Ge_nI_{n+1}Cl_{2n}$ ($n = 1$) to improve the optoelectronic properties by tuning the splitting energy $\Delta c$. The optimization window of optoelectronic properties prefers $\Delta c < 0$ eV because when the value of $\Delta c$ is larger than 0 eV, despite the reduced band gap and enhanced absorption, the relatively localized Ge $p_z$ orbital-derived band in the out-of-plane direction constitutes a new CBM with less dispersive, resulting in a relatively large $m_e$, which is not favored for photovoltaic applications. To achieve degenerate X($p_{x,y}$) and X($p_z$) bands ($\Delta c = 0$ eV), 9.35% in-plane biaxial compressive strain is needed, which is challenge to obtain in the present epitaxial strain engineering. In order to make this $\Delta c$ tuning more practical, next, we will explore the impact of the number of layers $n$ on the strain window for optoelectronic properties optimization, because the high-performance layered perovskite solar cells reported in the current experiments are usually based on $n = 3$ perovskite absorbers (e.g., $(BA)_2(MA)_2Pb_3I_{10}$, BA = $CH_3(CH_2)_3NH_3$).[8]

Taking $Cs_{n+1}Ge_nI_{n+1}Cl_{2n}$ with $n = 2$ and 3 (see Figure 5a) as an example, we plotted the splitting energy $\Delta c$ as a function of biaxial strain $\varepsilon$ in the representative layered perovskites, as shown in Figure 5b. It can be seen that a linear correlation between $\Delta c$ and $\varepsilon$ is observed in both cases, similar to the layered $CaAl_2Si_2$-type Zintl compounds.[13] This is because as the number of layers $n$ increases, the out-of-plane Ge-I bond length tends to show a linear relationship with strain changes (see Figure S4). For $Cs_3Ge_2I_3Cl_4$ and $Cs_4Ge_3I_4Cl_6$, the critical biaxial compressive strain reaching nearly zero $\Delta c$ value is 6% and 3.4%, respectively, as shown from the calculated orbital-projected band structure (see Figure 5c and 5d). Therefore, as $n$ increases (e.g., from 1 to 2 to 3), the strain window range for optoelectronic properties optimization will be greatly reduced (e.g., from 9.35% to 6% to 3.4%). In particular, for layered perovskites with $n \geq 3$, such critical strain may be experimentally achieved through lattice mismatch between the substrate materials and the perovskite films. This indicates that our orbital engineering strategy of tuning $\Delta c$ is practical to optimize the photovoltaic performance.

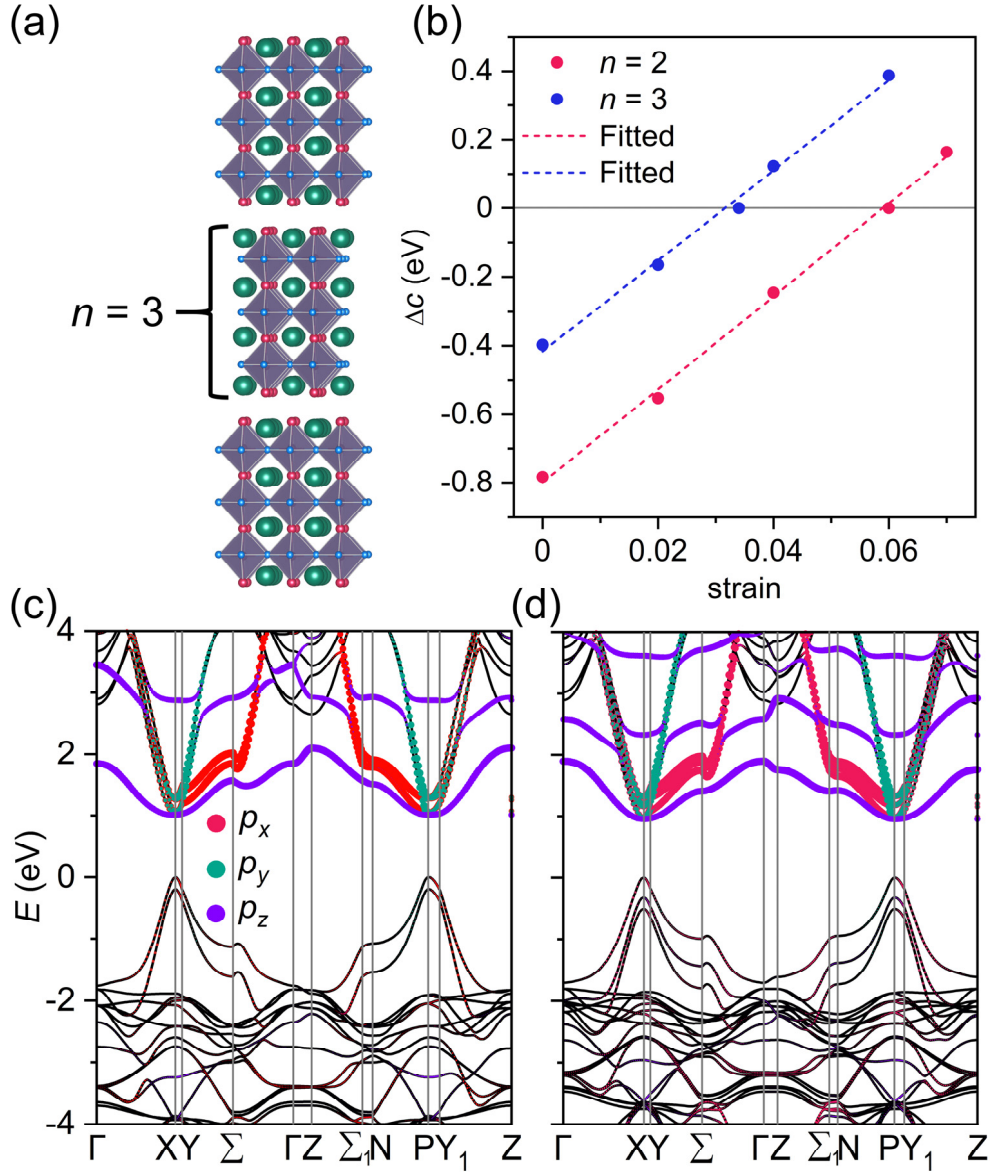

**Figure 5.** (a) Crystal structure of $Cs_4Ge_3I_4Cl_6$. (b) Splitting energy $\Delta c$ as a function of biaxial compressive strain $\varepsilon$ in the layered perovskites $Cs_{n+1}Ge_nI_{n+1}Cl_{2n}$ with $n = 2$ and 3. Orbital-projected band structure for (c) $Cs_3Ge_2I_3Cl_4$ and (d) $Cs_4Ge_3I_4Cl_6$ with $\Delta c = 0$ eV. $p_{x,y}$ and $p_z$ orbitals of $Ge^{2+}$ cations are projected on the band structure.

It is worth noting that, in addition to strain, other methods such as chemical doping can also be utilized to implement the orbital engineering approach, that is, manipulating the splitting of orbital energies to establish the orbital-property relationship in layered compounds. Moreover, the orbital engineering strategy proposed here can also be extended to other layered (-like) compounds with similar $p$ orbitals characteristics at the conduction or valence band edges. Figure S8 shows the orbital-projected band structures of $CsPb_2Br_5$ (space group: $I4/mcm$)[28], $Cs_3Sb_2I_9$ (space group: $P\bar{3}m1$)[29] and $CsPbI_3$ (space group: $P4/mbm$)[30]. All of these non-cubic halide perovskites possess $p$ orbital splitting characteristics at the conduction band edges, similar to the RP phase halide perovskite $Cs_2GeI_2Cl_2$. Therefore, it is desirable to establish similar orbital-property relationships to achieve the optimal optoelectronic performance of these materials.

## 3. Conclusions

In conclusion, we have employed $Cs_2GeI_2Cl_2$ as a representative example to demonstrate how to optimize the optoelectronic properties of layered (-like) halide perovskites through a simple band structure parameter based on first-principles method. Based on the calculated results, the orbital-property relationship based on $\Delta c$, which is defined by the splitting energy of $p$ orbitals of metal atom from the conduction band, is established as a guideline to optimize optoelectronic properties (e.g., band gap, transition dipole matrix elements, carrier effective masses, band width). Our results show that reducing the splitting energy $|\Delta c|$ can effectively reduce the band gap and carrier effective masses, while greatly improving the optical absorption that mainly resulting from the increased transition dipole matrix elements. Meanwhile, in order to obtain the best performance of layered perovskite with $n \geq 3$, the values of $\Delta c$ needs to be carefully controlled so as not to become a positive value, because the less dispersed $p_z$-derized CBM will lead to unfavorable electronic effective masses. Finally, the approach can be reasonably extended to several other non-cubic halide perovskites (e.g., $Cs_2PbBr_5$ and $Cs_3Sb_2I_9$) with similar $p$ orbitals characteristics at the conduction band edges. The orbital engineering approach presented in our work thus provides atomic-level guidance for understanding and optimizing the device performance of layered perovskite solar cells for experiments in the future.

## AUTHOR INFORMATION

**Corresponding Author**
*E-mail: hongjw@bit.edu.cn.
**Notes**
The authors declare no competing financial interests.

## ACKNOWLEDGMENT

This work is supported by the National Science Foundation of China (Grant No. 11572040), the Beijing Natural Science Foundation (Grant No. Z190011), and Graduate Technological Innovation Project of Beijing Institute of Technology. Theoretical calculations were performed using resources of the National Supercomputer Centre in Guangzhou, which is supported by Special Program for Applied Research on Super Computation of the NSFC-Guangdong Joint Fund (the second phase) under Grant No. U1501501.

**Supporting Information Available:** Computational details, calculation methods are available in the Supporting Information.